\documentclass{ws-ijtaf}[twocolumn,draft]

\usepackage{bm}
\usepackage{bbm}
\usepackage{listings}
\usepackage{subfigure}
\usepackage{amsmath}
\usepackage{amssymb}
\usepackage{booktabs}
\usepackage{hyperref}
\usepackage[T1]{fontenc}
\usepackage{epstopdf}

\hypersetup{colorlinks,citecolor=blue}

\newcommand*\diff{\mathop{}\!\kern0pt\mathrm{d}}

\begin{document}
\markboth{Fabien Le Floc'h}
{Pricing American options with the Runge-Kutta-Legendre finite difference scheme}

\catchline{}{}{}{}{}

\title{\uppercase{Pricing American options with the Runge-Kutta-Legendre finite difference scheme}}

\author{FABIEN LE FLOC'H}

\address{Delft Institute of Applied Mathematics, TU Delft\\ Delft, The Netherlands\\
	\email{fabien@2ipi.com } }

\maketitle

\begin{history}
	\received{(Day Month Year)}
	\revised{(Day Month Year)}
\end{history}

\begin{abstract}
This paper presents the Runge-Kutta-Legendre finite difference scheme, allowing for an additional shift in its polynomial representation.  A short presentation of the stability region, comparatively to the Runge-Kutta-Chebyshev scheme follows. We then explore the problem of pricing American options with the Runge-Kutta-Legendre scheme under the one factor Black-Scholes and the two factor Heston stochastic volatility models, as well as the pricing of butterfly spread and digital options under the uncertain volatility model, where a Hamilton-Jacobi-Bellman partial differential equation needs to be solved. We explore the order of convergence in these problems, as well as the option greeks stability, compared to the literature and popular schemes such as Crank-Nicolson, with Rannacher time-stepping.
\end{abstract}


\section{Introduction}




Applied to the diffusion partial differential equation (PDE), which arises in many financial models (the option pricing model of \citet{black1973pricing}, the local volatility model of \citet{dupire1994pricing}, the stochastic volatility model of \citet{heston93}), the explicit Euler finite difference method is very simple to implement. Being explicit, it does not require the solution of any linear system. Furthermore, with a sparse system, the matrix-vector multiplications involved are fast. The explicit Euler method is however only first-order in time. The use of a predictor-corrector scheme, or of the second-order explicit Runge-Kutta scheme (RK2) make it possible to reach second-order convergence in time. The bigger issue is the limitation on the time-step size imposed by the stability condition. The latter implies that time-step size $\Delta t$ follows a square relationship with the asset step size $\Delta x$ (and variance step size $\Delta v$ for stochastic volatility models).
For practical grid geometries, more than 10 000 time-steps may be required for the explicit scheme to be stable.

Super time-stepping methods, typically based on the boundedness of Chebyshev polynomial, allow us to circumvent the time-step size limitation for parabolic (mildly stiff) problems. \citet{OSullivan09} applied the super time-stepping of \citet{alexiades1996super} to the pricing of American options in the Black-Scholes model, using a Richardson extrapolation to attain second-order accuracy. \citet{in2010adi} applied the second-order Runge-Kutta-Chebyshev solver of \citet{sommeijer1998rkc} to compute the price of European and barrier options under the Heston model. Here, we consider a more recent development with increased stability properties due to \citet{meyer2014stabilized}, where a Legendre polynomial defines the recursion of the super time-stepping method, leading to the Runge-Kutta-Legendre (RKL) scheme. Their approach does not require us to define and guess an extra damping parameter and is directly second-order.

This paper focuses on the solution of the linear complementary problem (LCP), which arises in the pricing of American options. As a main contribution, we explain how to apply the RKL scheme  to a variety  of PDEs arising in finance, which include non-linearities and/or multiple dimensions. We thus show that the RKL scheme is an interesting alternative to more classical implicit finite difference schemes on many of those problems, with the main advantage of being straightforward to implement. We also present some limitations of the scheme on specific examples, where some additional damping may be necessary in the RKL scheme.

 In Section 2, we introduce the RKL scheme. We allow for a non-zero shift in the polynomial representation in similar fashion as \citet{verwer1999second} does in the context of the Runge-Kutta-Chebyshev scheme and study the stability regions of the schemes. This shift allows us to increase the damping properties of the scheme when necessary. In Section 3, we explain how to discretize the Black-Scholes or the Dupire local-volatility PDE with the RKL scheme and we detail how to handle the LCP. In Section 4, we measure the order of convergence of the scheme on American options under the Black-Scholes model, and take a look at the smoothness of the option greeks. In Sections 5 and 6, we apply the scheme respectively to the Hamilton-Jacobi-Bellman PDE of the uncertain volatility model of \citet{avellaneda1995pricing}, and to American options under the Heston stochastic volatility model. We look at the convergence of the RKL scheme on such problems.

\section{The Runge-Kutta-Legendre scheme}
\subsection{Overview}
For a time discretization defined by $(t_j)_{j \in \{0,..,n\}}  ~~,~~ k_j = t_j-t_{j-1}$ where $t_0$ is typically the valuation time and $t_n$ the option expiry, the Runge-Kutta-Legendre scheme of order-2 (RKL) consists of $s$ explicit 
stages to define the value at $t_{j-1}$ from an initial value at $t_j$.

Let $\mathcal{L}$ be the Black-Scholes-Merton operator defined by:
\begin{equation}
\mathcal{L}\left(f(x,t),x,t\right) =  - \frac{1}{2}\sigma(x,t)^2 x^2\frac{\partial^2 f }{\partial x^2}  - \mu(x,t) x \frac{\partial f}{\partial x}+ r(x,t)f(x,t)\,,
\end{equation}
where $x$ is the underlying price, $\mu$ is the underlying drift, $\sigma$ its volatility and $r$ the interest rate, and $F(x)=f(x,t_n)$ the option payoff at maturity.
With this notation, the Black-Scholes-Merton equation is
\begin{equation}
\frac{\partial f}{\partial t}(x,t) = \mathcal{L}\left(f(x,t),x,t\right) \,.
\end{equation}
And the RKL scheme reads \citep{meyer2014stabilized}
\begin{subequations}
	\begin{align}
	\hat{f}^0 &= f(t_{j})\,,\\\label{eqn:rkl2_stage0}
	\hat{f}^1 &=  \hat{f}^0 +  \tilde{\lambda}_1 k_j \mathcal{L} \left(\hat{f}^0\right)\,,\\\label{eqn:rkl2_stage1}
	\hat{f}^{\eta} &= \lambda_\eta \hat{f}^{\eta-1} + \nu_\eta \hat{f}^{\eta-2} + (1-\lambda_\eta -\nu_\eta)\hat{f}^0 +  \tilde{\lambda}_\eta k_j \mathcal{L} \left(\hat{f}^{\eta-1}\right) + \tilde{\gamma}_\eta k_j \mathcal{L} \left(\hat{f}^{0}\right) \,,\nonumber\\&\quad \textmd{ for } 2 \leq \eta \leq s\,,\\
	f(t_{j-1}) &= \hat{f}^s\,,\label{eqn:rkl2_stages}
	\end{align}
\end{subequations}
with the parameters
\begin{align*}
b_{\eta-1} = \frac{P_{\eta}''(w_0)}{P_\eta'(w_0)^2}\,,&\quad a_{\eta-1} = 1 - b_{\eta-1}P_\eta(w_0)\,,\\
\lambda_\eta =\frac{2\eta-1}{\eta}\frac{b_{\eta-1}}{b_{\eta-2}}w_0\,,&\quad
\tilde{\lambda}_\eta=  \frac{2\eta-1}{\eta}\frac{b_{\eta-1}}{b_{\eta-2}} w_1\,,\\
\nu_\eta =-\frac{\eta-1}{\eta}\frac{b_{\eta-1}}{b_{\eta-3}}\,,&\quad \tilde{\gamma}_\eta = - a_{\eta-2} \tilde{\lambda}_\eta\,,
\end{align*}
for $2 \leq \eta \leq s$ and $b_0 = b_1$, $a_0=1-b_0 w_0$, $\tilde{\lambda}_1 = b_0 w_1$, $\nu_2 = - b_1 / (2b_0)$.
We generalized the method to shifted Legendre polynomials and allow for a non-zero shift $o=\frac{\epsilon}{s^2}$ in our formulation, in a similar fashion as \citet{verwer1999second} does for Chebyshev polynomials. Then we have
\begin{equation}
w_0=1+o =1 + \frac{\epsilon}{s^2}\,,\quad w_1 = \frac{P_s'(w_0)}{P_s''(w_0)}\,,
\end{equation}
where $P_s$ is the Legendre polynomial of degree $s$ defined by the recursion
\begin{equation*}
\eta P_{\eta}(w) = (2\eta - 1) w P_{\eta-1}(w) - (\eta - 1) P_{\eta-2}(w)\,,\quad \textmd{for } \eta= 2,...,s\,,
\end{equation*}
with $P_0=1$, $P_1(w)=w$. The first and second derivatives may be computed in the same recursion by applying the chain rule of differentiation.

In contrast with the Runge-Kutta-Chebyshev method, the RKL method is already stable without shift, when $\epsilon=0$, which is the recommended value in \citep{meyer2014stabilized}. Without shift, we have then \begin{equation*}w_1 = \frac{4}{s^2+s-2}\,,\quad b_{\eta-1} = \frac{\eta^2 + \eta - 2}{2\eta(\eta+1)}\,,\quad a_{\eta}=1-b_\eta\,.\end{equation*}
The scheme is stable as long as \begin{equation}
k_j \leq \frac{1+w_0 }{2w_1}\Delta t_{\textsf{explicit}} \label{eqn:rkl_timestep}
\end{equation} where $\Delta t_{\textsf{explicit}}$ is the maximum time-step size allowed by the explicit Euler method.


Equation \ref{eqn:rkl_timestep} gives us a method to automatically determine the minimum number of stages $s$ for a given time-step size $k_j$. Without shift, the solution is explicit. With a shift, we may solve for the condition iteratively. The number of stages $s$ is proportional to the inverse square root of $\Delta t_{\textsf{explicit}}$, and thus, the RKL scheme should be more efficient than the classic explicit Euler scheme. In our numerical examples, we will take the first odd integer larger than, or equal to, the minimum number of stages allowed.

\subsection{Stability region}
For a finite difference method, absolute stability is defined using the simple problem:
\begin{align}\label{simple_problem}
u'(t) &= \lambda u(t)
\end{align} 
where $\lambda$ is a complex number. The application of a numerical scheme to this problem will lead to a condition on $z = k \lambda$, where $k$ is the time step, in order to ensure convergence \citep{Le07}. For the forward Euler scheme, the discretization leads to $u_{j+1} = (1+k \lambda) u_{j}$, its stability region will be defined by ${\lvert 1+z \lvert} \leq 1$.

\begin{figure}[h]
	\centering{
		\subfigure[\label{fig:leg_cheb_21s_rkl}RKL $s=21$ stages  with $\epsilon=0$ and $\epsilon=20$.]{
			\includegraphics[width=.45\textwidth]{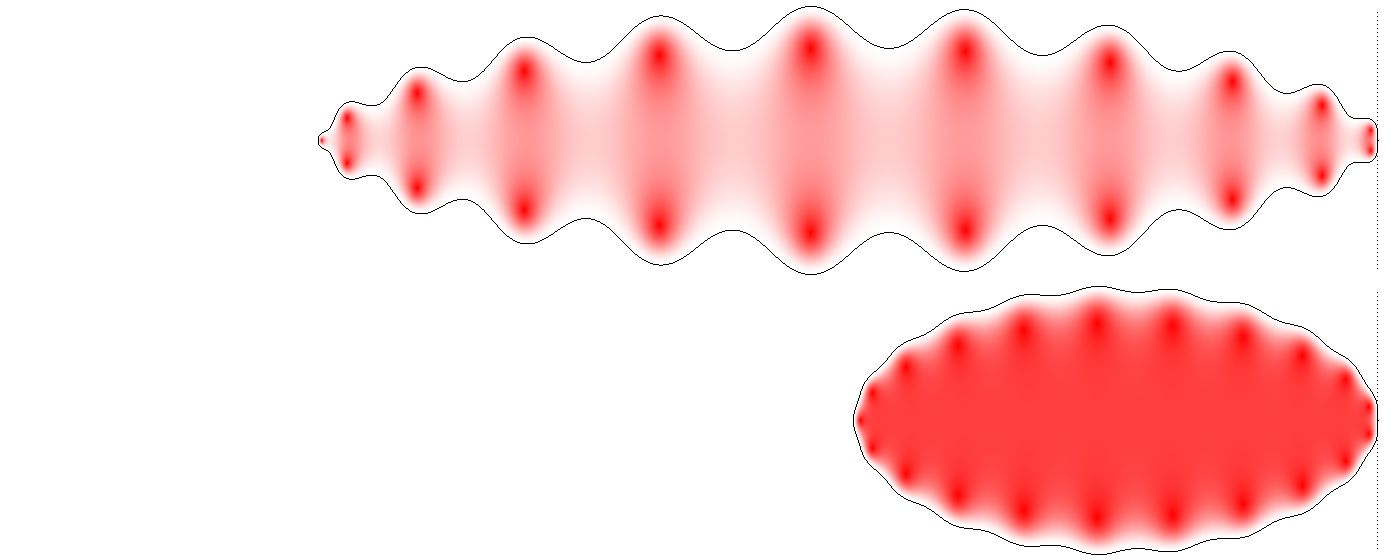}}
		\subfigure[\label{fig:leg_cheb_11s_rkl}RKL $s=11$ stages with $\epsilon=0$ and $\epsilon=20$.]{
			\includegraphics[width=.45\textwidth]{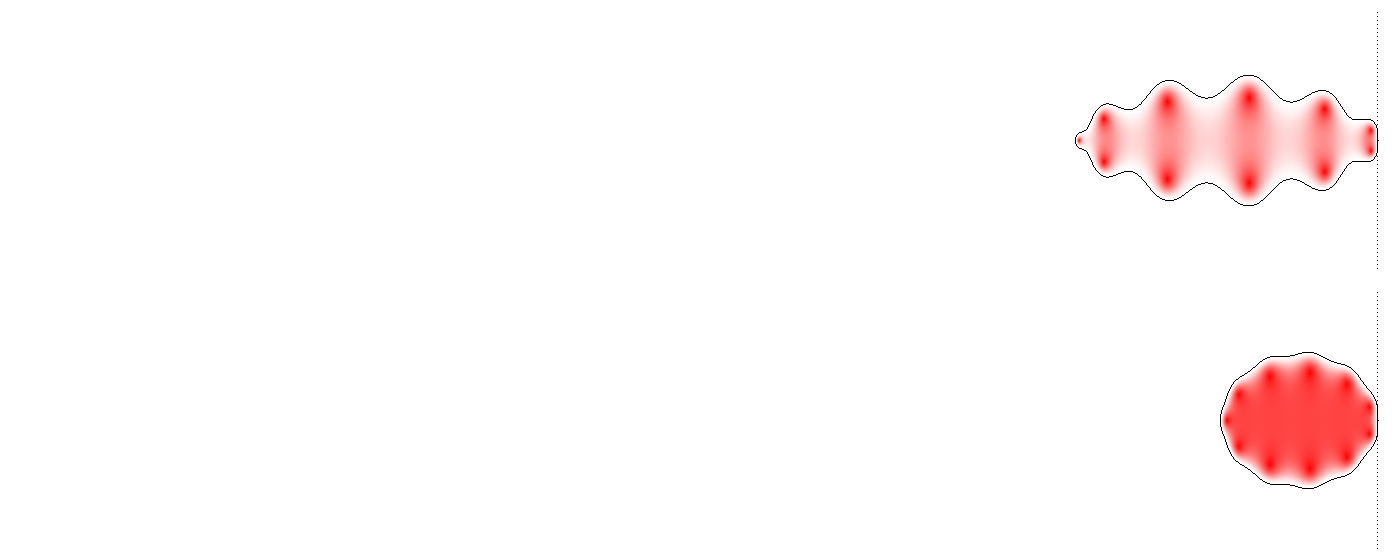} }\\
		\subfigure[\label{fig:leg_cheb_21s_rkc}RKC $s=21$ stages, $\epsilon=2.3$, $\epsilon=23$, $\epsilon=0.01$.]{
			\includegraphics[width=.45\textwidth]{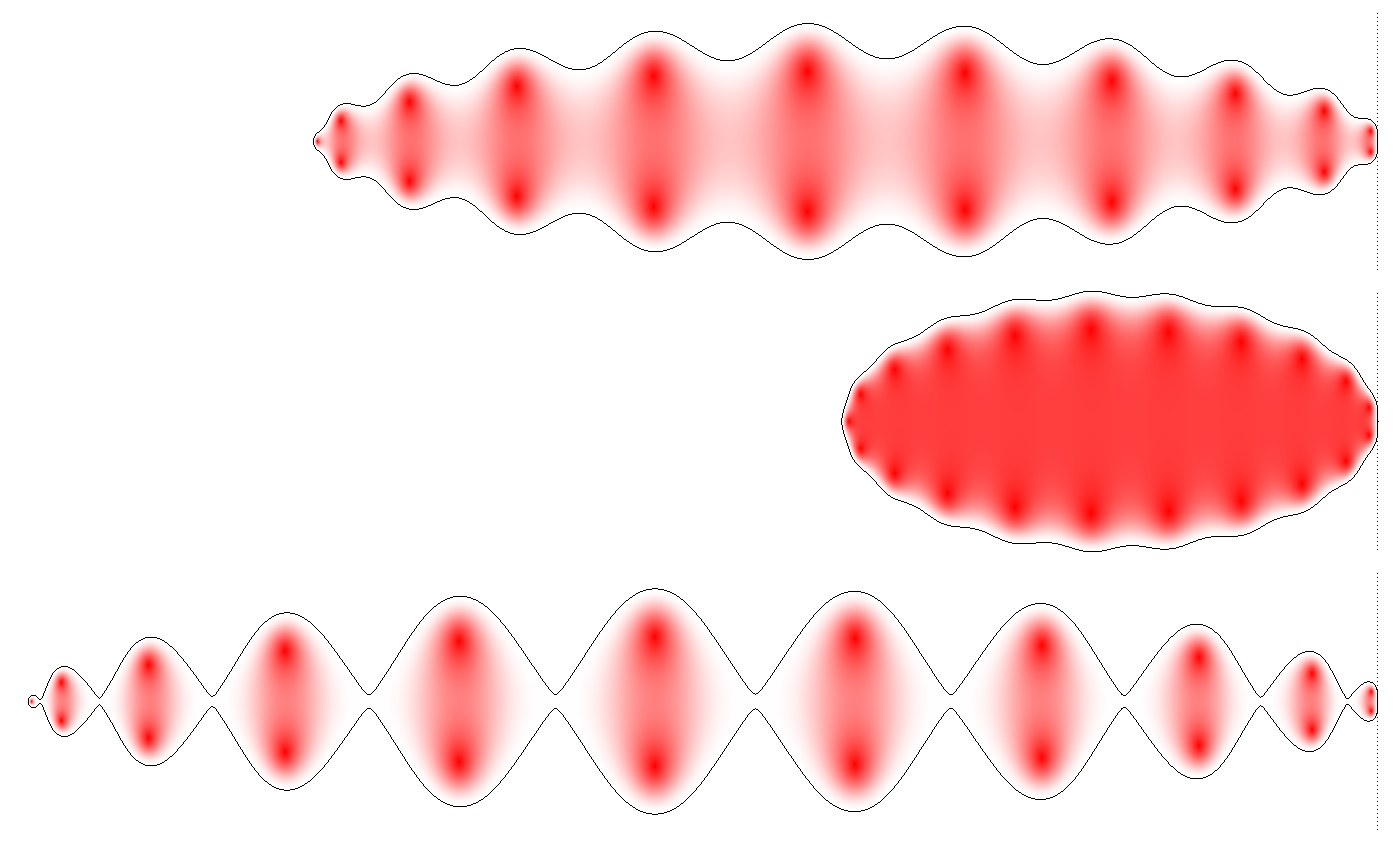}}
			\subfigure[\label{fig:leg_cheb_11s_rkc}RKC $s=11$ stages, $\epsilon=2.3$, $\epsilon=23$, $\epsilon=0.01$.]{
			\includegraphics[width=.45\textwidth]{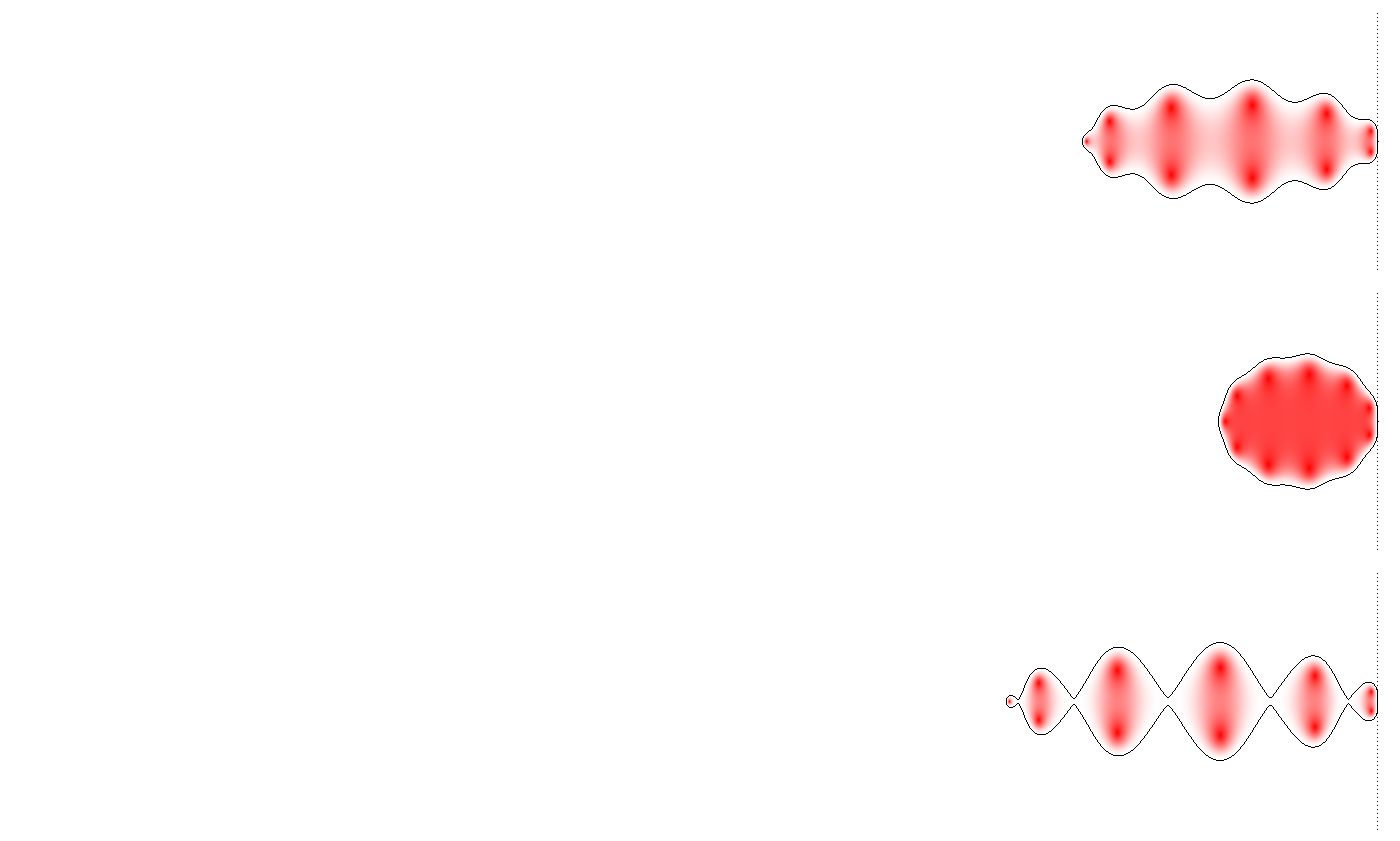} }\\
		
		\caption{Stability regions. The origin is located on the right boundary, in the middle of each graph. RKL stands for the Runge-Kutta-Legendre scheme and RKC for the Runge-Kutta-Chebyshev scheme.}}
\end{figure}

In Figures \ref{fig:leg_cheb_21s_rkl}, \ref{fig:leg_cheb_21s_rkc}, \ref{fig:leg_cheb_11s_rkl}, \ref{fig:leg_cheb_11s_rkc}, we plot the stability region of the Runge-Kutta-Legendre scheme using $s=21$ stages and $s=11$ stages. We color the stability region proportionally to the damping factor $\left|{u_{j+1}}/{u_{j}}\right|$. If the damping factor is close to 1, there is almost no damping, and the color is close to white. If the damping factor is close to 0, there is strong damping, and the color is close to deep red. We consider the case of a non-shifted Legendre polynomial and a shift $\epsilon=20$.

We also plot the stability region of the Runge-Kutta-Chebyshev scheme, where we adjust the shift such that the maximum time-step size allowed for stability is the same for RKC and RKL. The last plot is a Runge-Kutta-Chebyshev stability region with a small shift $\epsilon=0.01$.

The stability region of the RKL scheme with $\epsilon=0$ is the largest in terms of area, even compared to the RKC scheme with a very small shift (Table \ref{tbl:stability_area}). The former is shorter along the real axis, but wider along the imaginary axis. Damping is also stronger on average, and is more uniformly distributed in the stability region. Furthermore, the stability region of the RKC scheme exhibits many regions of very short height along the imaginary axis, which could easily become an issue for problems involving  small imaginary eigenvalues (related to convection for example). The stability region of the RKC scheme is only comparable with a relatively large shift $\epsilon=2.3$ (although it is still narrower along the imaginary axis). With an even larger shift $\epsilon \approx 20$, the stability region of the two schemes become much more similar. Looking at the stability region, without taking into consideration the damping factor inside the stability region may be misleading. Beside an increase of the width of the stability region along the imaginary axis relatively to the real axis, the more important property of the shift is to increase the damping significantly (or equivalently reduce the average damping factor). In Figures \ref{fig:leg_cheb_21s_rkl} and \ref{fig:leg_cheb_21s_rkc}, the stability region is much darker with no pale spots.
\begin{table}[h]
	\caption{The area of the stability region is expressed as a percentage of the area of the stability region of the RKL scheme with $\epsilon=0$. Damping is the average damping factor inside the stability region.\label{tbl:stability_area}}
	\centering{
		\begin{tabular}{lrrrrrr}\toprule
			Scheme &	 \multicolumn{2}{c}{$s=21$} & \multicolumn{2}{c}{$s=11$} & \multicolumn{2}{c}{$s=7$}\\\cmidrule(lr){2-3}\cmidrule(lr){4-5}\cmidrule(lr){6-7}
			& Area & Damping & Area & Damping & Area & Damping\\
	
RKC, $\epsilon=0.01$ & 90\% & 0.60 & 92\% & 0.61 &  95\% &0.61 \\
RKL, $\epsilon=0$ & 100\% & 0.55  & 100\% & 0.56 & 100.0\% &0.57\\
RKC, $\epsilon=2.3$ & 85\% & 0.51& 88\% & 0.51 &93\% &0.52  \\
RKL, $\epsilon=20$ & 27\% & 0.25 & 33\% & 0.28&43\% &0.33 \\
RKC, $\epsilon=23$ & 26\% & 0.25 & 33\% & 0.28&42\% &0.33 \\\bottomrule
	\end{tabular}}
\end{table}

To conclude this section, the RKL scheme without shift provides already a practical stability region. A main advantage of the RKL scheme over the RKC scheme is however the Convex Monotone Property (CMP) \citep{meyer2014stabilized}, which we will describe in more details in Section \ref{sec:cmp}.

\section{Discretization of the Black-Scholes or local volatility PDE}
For a space discretization defined by $(x_i)_{i \in \{0,..,m\}}  ~~,~~ h_i = x_i-x_{i-1}$
The central difference operator $\mathcal{D}_x$ is defined as
\begin{equation}
\mathcal{D}_x f_{i,j} =\frac{h_i^2 f_{i+1,j} + \left(h_{i+1}^2-h_i^2\right)f_{i,j} - h_{i+1}^2f_{i-1,j}}{h_i h_{i+1}\left(h_{i+1} + h_i\right)}\,.
\end{equation}

The central second difference operator $\mathcal{D}_x^2$ is defined as
\begin{equation}
\mathcal{D}_x^2 f_{i,j} =  2 \frac{h_i f_{i+1,j} - \left(h_{i+1}+h_i\right)f_{i,j} + h_{i+1}f_{i-1,j}}{h_i h_{i+1}\left(h_{i+1} + h_i\right)}\,.\label{eqn:second_derivative}
\end{equation}

For $i \in \{1,...,m-1\}$, $j \in \{1,...,n\}$, let $\mathcal{L}_{i,j}=(r_{i,j}I-\mu_{i,j} x_i \mathcal{D}_{x}-\frac{1}{2}\sigma_{i,j}^{2} x_i^2\mathcal{D}_{x}^{2})$ be the discrete differential operator where $\mu_{i,j}$ is the drift, $\sigma_{i,j}$ the volatility and $r_{i,j}$ the interest rate on the interval $[t_{j-1},t_j]$. 
The first explicit stage  corresponding to Equation  \ref{eqn:rkl2_stage1}  reads
\begin{equation}
\hat{f}^1_{i,j} = f_{i,j} - k_j \bar{\lambda}_1 \mathcal{L}_{i,j}f_{i,j}\,.
\end{equation}
According to Equation \ref{eqn:rkl2_stages}, the other stages involve the explicit scheme on the time-step of size $k_j \bar{\lambda}_\eta$. Each stage thus consists in a single matrix vector multiplication, where the matrix is tridiagonal with main diagonal $(b)_{i=0,...,m}$, lower diagonal $(a)_{i=1,...,m}$ and upper diagonal $(c)_{i=0,...,m-1}$ such that
\begin{equation*}
\hat{f}^\eta_{i,j} = \hat{g}^{\eta-1}_{i,j} +\bar{\lambda}_\eta \left( a_{i,j}\hat{f}^{\eta-1}_{i-1,j} +  b_{i,j} \hat{f}^{\eta-1}_{i,j} +  c_{i,j}\hat{f}^{\eta-1}_{i+1,j} \right)\,,
\end{equation*}
with
\begin{subequations}
\begin{align}
a_{i,j} & = -\frac{k_{j}}{h_i\left(h_{i+1}+h_i\right)}\left(\mu_{i,j} h_{i+1} x_i - \sigma_{i,j}^2 x_i^2\right) \label{tridiagonal_tr_eq1}\,,\\
b_{i,j} & = - k_{j} \hat{b}_{i,j}\,,\quad \hat{b}_{i,j}=\left(r_{i,j}+ \frac{\mu_{i,j}(h_i-h_{i+1})+\sigma_{i,j}^2 x_i^2 }{h_i h_{i+1}} \right) \label{tridiagonal_tr_eq2}\,,\\
c_{i,j} & = \frac{k_{j}  }{h_{i+1}\left(h_{i+1}+h_i\right)}\left(\mu_{i,j}h_i x_i+ \sigma_{i,j}^2 x_i^2\right) \label{tridiagonal_tr_eq3}\,,
\end{align}
\end{subequations}
and
\begin{align}
 \hat{g}^{\eta-1}_{i,j} = \lambda_\eta \hat{f}^{\eta-1}_{i,j} + \nu_\eta \hat{f}^{\eta-2}_{i,j} + (1-\lambda_\eta -\nu_\eta)\hat{f}^0_{i,j} + \frac{\tilde{\gamma}_\eta}{\bar{\lambda}_1}  \left(\hat{f}^1_{i,j} - \hat{f}^{0}_{i,j}\right) \,,\quad \textmd{ for } 2 \leq \eta \leq s\,,
\end{align}
and $\hat{g}^0_{i,j} = \hat{f}^0_{i,j}$.
The coefficients are kept constant during the full time-step $k_j$.

With this discretization, the maximum allowed time-step size for the explicit Euler scheme reads $\Delta t_{\textsf{explicit}} = \frac{1}{\max(\hat{b}_{i,j})}$.
\subsection{Boundary conditions}
We consider here the boundary condition where we assume that $\frac{\partial^2 f }{\partial x^2} = 0$ at the boundaries.
This is true for all payoffs linear at the boundaries, which is a reasonable assumption for most payoffs \citep{windcliff2004analysis}.
The Black-Scholes equation becomes:
\begin{equation}
\frac{\partial f}{\partial t}(x,t) + \mu(x,t)x\frac{\partial f}{\partial x}(x,t) = r(x,t)f(x,t)
\end{equation}

We discretize the derivative by a first order approximation in the variable $x$. 
This is reasonable because the first order error in $x$ is proportional to the Gamma, which we assumed to be $0$:
\begin{equation}
\mathcal{D}^+_x f_{i,j} = \frac{f_{i+1,j}-f_{i,j}}{h_{i+1}}\,.
\end{equation}
This leads to 
\begin{align}
{b}_{0,j} = - k_{j}\left(r_{0,j}+\frac{\mu_{0,j}x_0}{h_1}\right)\,,\quad & {c}_{0,j} = k_{j} \frac{\mu_{0,j}x_0}{h_1}\,,
\end{align}
for the lower boundary and 
\begin{align}
{a}_{m,j}  = -\alpha k_{j}\frac{\mu_{m,j}x_m}{h_m}\,,\quad & {b}_{m,j} = - k_{j}\left(r_{m,j}-\frac{\mu_{m,j}x_m}{h_m}\right)\,,
\end{align}
for the upper boundary.

\subsection{Illustration of the Convex Monotone Property}\label{sec:cmp}
The Convex Monotone Property (CMP) is defined for a parabolic PDE with variable diffusion coefficient ($\sigma^2 x^2$ in the Black-Scholes and local volatility PDEs). When the initial condition is monotone, the CMP ensures that the solution will not only
be monotone for a single value of the diffusion coefficient, but will also maintain the monotonicity property for any diffusion coefficient between
zero and the maximal value allowed for stability, a convex set of values \citep{meyer2014stabilized}. This is a desirable property for non-linear problems, where the non-linearity has some dependency onto the monotonicity of the solution at each time-step. A non-monotonicity preserving scheme may fail to converge or converge to the wrong solution on those problems.

\begin{figure}[!h]
	\begin{center}
		\includegraphics[width=\textwidth]{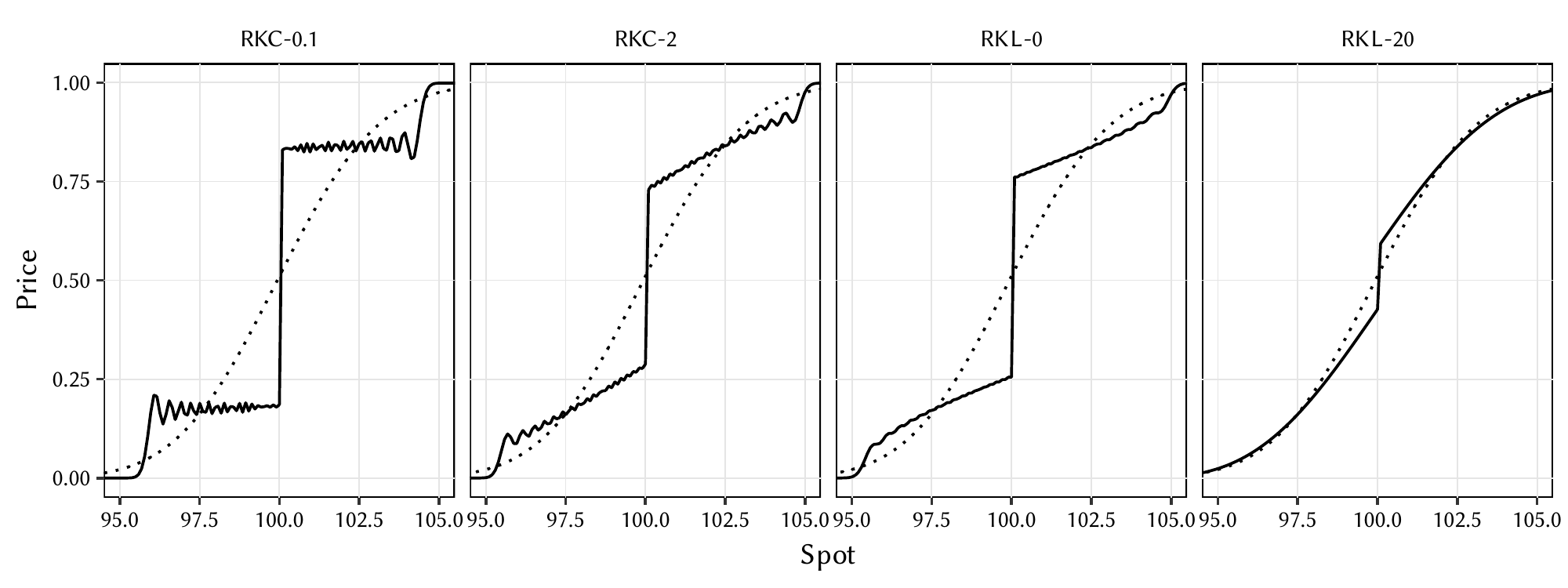}
		\caption{Value of a digital option after one time-step for different finite difference schemes. RKC-0.1 and RKC-2 correspond to the Runge-Kutta-Chebyshev scheme with shift $\epsilon=0.1$ and $\epsilon=2$. RKL-0 and RKL-20 are the Runge-Kutta-Legendre schemes with shift $\epsilon=0$ and $\epsilon=20$. The dotted line indicates the analytical value.}
		\label{fig:digital_1step}
	\end{center}
\end{figure}

We consider a binary call option of strike $K=100$, paying \$ $1$ at maturity $T=0.25$ with volatility $\sigma=25\%$, interest rate $r=10\%$, on a uniform grid composed of 800 steps in the asset price dimension starting at $x_{\min} = 68.71$ up to $x_{\max}=145.58$ (which correspond to three standard deviations). Figure \ref{fig:digital_1step} displays the value of the option in the finite difference grid after one time-step of size $k_n=0.01$, at $t=T-k_n$.
The RKL scheme without shift requires then $s=79$ stages for stability, and $s=111$ stages with shift $\epsilon=20$. In Figure \ref{fig:digital_1step}, we use $s=111$ stages for both cases such that the only difference is the shift size. We noticed that the number of stages did not impact the values significantly: the price plot is essentially the same with $\epsilon=0$ and $s=79$ or $s=111$ (or even with $s=790$).

Small oscillations are present when the value is computed by the Runge-Kutta-Chebyshev scheme with shift $\epsilon=0.1$ or $\epsilon=2$, and the option value is non-monotonic. With the RKL scheme, the option value stays monotonic, even without shift, because of the monotone convex property of the scheme.
A larger shift $\epsilon=20$ leads to a smoother, and more accurate value. 

Due to the explicit nature of the schemes, the option value is not so smooth after one time-step, which may be an issue for the option greeks. It will however become smooth after several time-steps. And thus we may also choose to use finer time-steps to smooth-out the solution. On this example, the solution is smooth after 10 time-steps of size 0.001 (instead of 1 time-step of size 0.01).

\subsection{American Option Specifics}
The early exercise feature of the option adds a free boundary on top of the Black-Scholes partial differential equation. Let $f$ be the option price, the following system of partial differential inequalities is satified \citep{LaLa96}:

\begin{equation}\label{lcp}
\left. \begin{gathered}
\frac{\partial f}{\partial t}(x,t) + \mu(x,t)x \frac{\partial f}{\partial x}(x,t)   + \frac{1}{2}\sigma(x,t)^2 x^2 \frac{\partial^2f }{\partial x^2}(x,t) \leq r(x,t)f(x,t)\\
\left(\frac{\partial f}{\partial t}(x,t) + \mu(x,t) x\frac{\partial f}{\partial x}(x,t) + \frac{1}{2}\sigma(x,t)^2 x^2 \frac{\partial^2f }{\partial x^2}(x,t) - r(x,t)f(x,t)\right)\left(f-F\right)=0\\
f \geq F
\end{gathered} \right\}
\qquad \text{}
\end{equation}

Let $M_j$ be the tridiagonal matrix with lower diagonal $a_{i,j}$ for $i \in \{1,...,m\}$, upper diagonal $c_{i,j}$ for $i \in \{0,...,m-1\}$ and diagonal $b_{i,j}$ for $i \in \{0,...,m\}$, with $a_{i,j}$, $b_{i,j}$, $c_{i,j}$ defined by Equations \ref{tridiagonal_tr_eq1}, \ref{tridiagonal_tr_eq2}, \ref{tridiagonal_tr_eq3}. Let $f_j = (f_{0,j},...,f_{m,j})$, $g_j = (g_{0,j},...,g_{m,j})$, we discretize the linear complimentary problem (\ref{lcp}) explicitly with the RKL scheme, stage by stage, for $j=n,...,1$:
\begin{subequations}
\begin{align}\label{eqn:rkl_american}
\hat{f}^1 &= \max\left(F, f_j +\tilde{\lambda}_1 k_j M_j f_j\right)\,, \\
\hat{f}^{\eta} &= \max\left(F, 
 \lambda_\eta \hat{f}^{\eta-1} + \nu_\eta \hat{f}^{\eta-2} + (1-\lambda_\eta -\nu_\eta)\hat{f}^0 + \right.\nonumber\\
& \left. \tilde{\lambda}_\eta k_j M_j \left(\hat{f}^{\eta-1}\right) + \tilde{\gamma}_\eta k_j M_j \left(\hat{f}^{0}\right) \right) \,,\quad \textmd{ for } 2 \leq \eta \leq s\,,\\
f_{j-1} &= \hat{f}^s\,.
\end{align}
\end{subequations}

\section{Convergence}
\subsection{Greeks stability}
We compute the $\Gamma$ (the second derivative of the option price) of 1 year American Put option of strike $K= 160$ with spot $S=100$, 40\% volatility , 5\% interest rate on a grid composed of 500 space steps and 80 time steps. The option is therefore in the money.
The $\Gamma$ is computed through the finite difference discretization of the second derivative (Equation \ref{eqn:second_derivative}). 

The same grid, and thus the same time-step size is used for all schemes considered, namely the standard Crank-Nicolson scheme, the Crank-Nicolson scheme where the first two time-steps are replaced by four half-steps of the implicit Euler scheme, known as Rannacher smoothing \cite{GiCa2006}, and the RKL scheme. The latter scheme makes use of 51 internal stages at each time-step, in relation with the stability condition (Equation \ref{eqn:rkl_timestep}). When compared to the Rannacher scheme with the optimized tridiagonal solver of \citet{brennan1977valuation} for the vanilla American option problem, under positive interest rates, the RKL scheme is around eight times slower. But if a more generic Projected Successive Over Relaxation solver (PSOR) \citep{WiDeHo93} is employed to solve the non-linearity, the Rannacher scheme is then more than 3 times slower than the RKL scheme on this problem (Table \ref{tbl:gamma_perf}). Contrary to the Brennan-Schwartz optimized tridiagonal solver, the PSOR solver is still applicable under negative interest rates, for non-monotonic option payoffs, or more sophisticated, non-tridiagonal, discretizations.

\begin{figure}[h]
	\begin{center}
		\includegraphics[width=\textwidth]{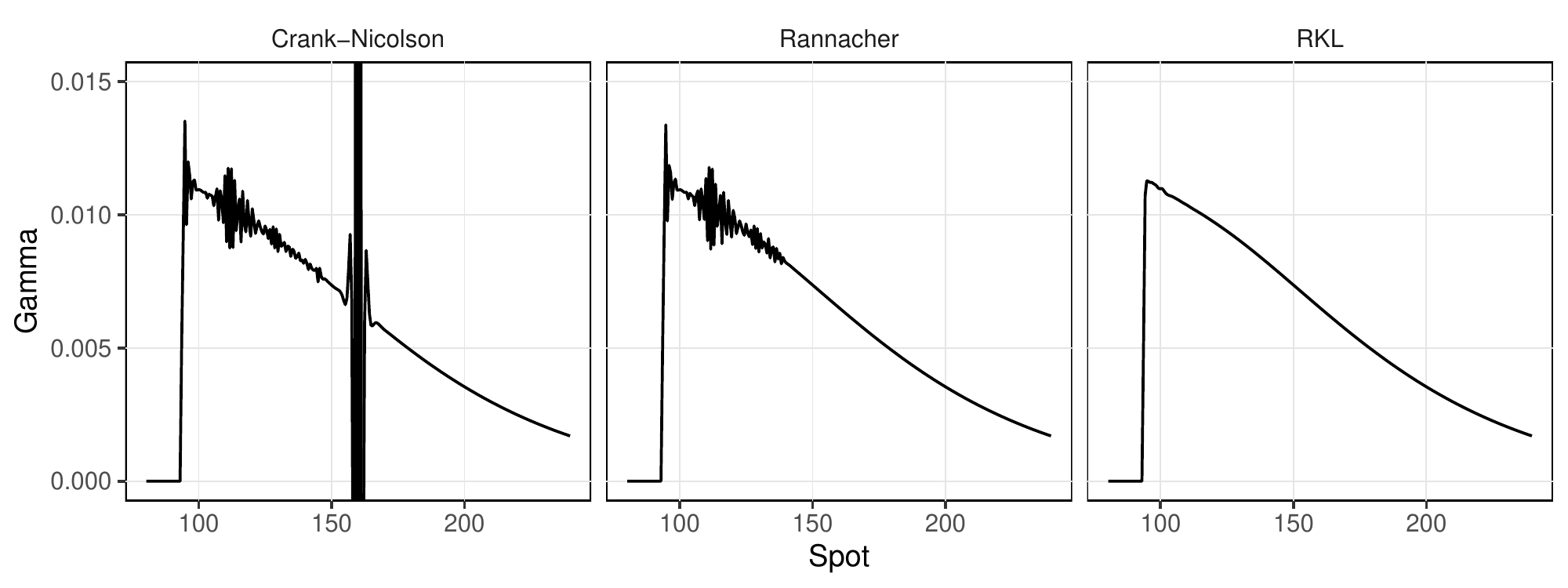}
		\caption{$\Gamma$ of an In-the-money American put Option with strike at 160 and 80 time steps.}
		\label{fig:gamma_put}
	\end{center}
\end{figure}

The $\Gamma$ with the RKL scheme is smooth while Crank-Nicolson presents oscillations at the payoff discontinuity. 
Rannacher smoothing works well for European options \citep{GiCa2006}: it produces smooth greeks and improves the convergence. But it is not true for American options as shown in Figures \ref{fig:gamma_put}.
If the Rannacher smoothing is only applied at maturity, some oscillations are left: the ones that corresponds to the early exercise boundary. In order to have smooth greeks for American options, one would need to add backward Euler steps after every potential discontinuity introduced by the early exercise feature. The problem is that the resulting scheme loses 1 order of accuracy \citep{lefloch2014tr}.

\begin{table}[h]
	\centering{
		\begin{tabular}{l r}\toprule
			Scheme & Time (ms)\\\midrule
			Rannacher with Brennan-Schwartz tridiagonal solver & 1.6 \\
			Rannacher with PSOR solver & 39.5 \\
			RKL &  12.2\\\bottomrule
		\end{tabular}
		\caption{Time in ms to compute the $\Gamma$ of an American option with 500 space-steps and 80 time-steps.\label{tbl:gamma_perf}}
	}
\end{table}

\subsection{Convergence for a fixed space step}
We follow \citet{OSullivan09} and look at the convergence of a 1 year American put option of strike price $K = 100$, using a spot price $S = 100$, a discount rate $r= 5\%$, and a volatility $\sigma=20\%$. We fix the space step size at 1.0 on a grid ranging from $S_{\min}=0$ to $S_{\max}=500$. This corresponds to 500 space steps. Note that this places the strike and the spot on the grid, which is important to avoid the indroduction of additional errors from the payoff discretization in the grid \citep{pooley2003convergence}.

We use the same theoretical value of 6.0874933186 as in their paper. Note that this is not the exact American option price because we have fixed the space step size. We compare the convergence of the Runge-Kutta-Legendre scheme (named "RKL"), Rannacher (named "RAN"), Crank-Nicolson with the Brennan-Schwartz solver (named "CN").

\begin{figure}[!h]
	\begin{center}
		\includegraphics[width=\textwidth]{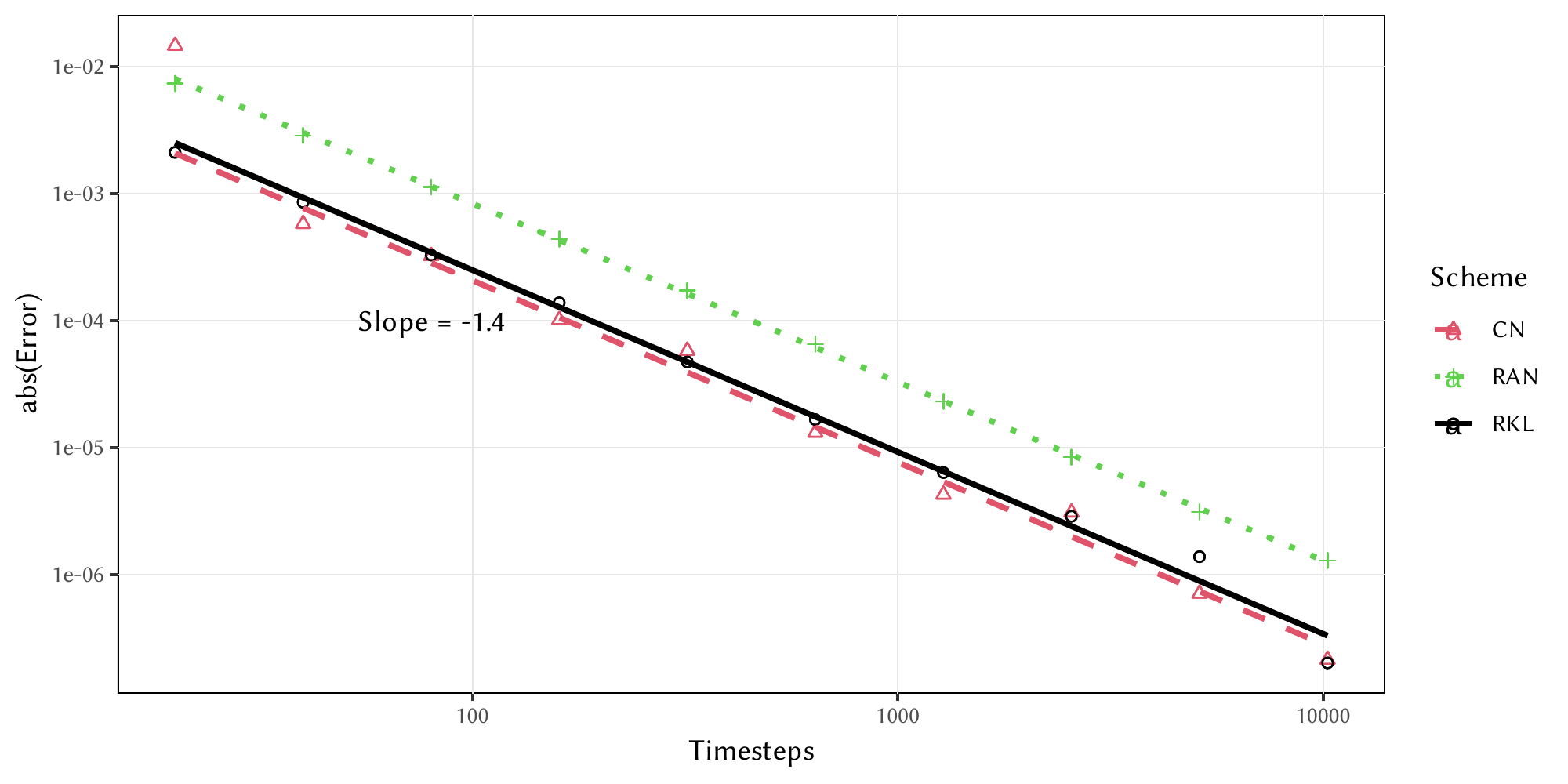}
		\caption{Convergence for an American Put with a fixed space step. Crank-Nicolson "CN", Rannacher "RAN" with the Brennan-Schwartz  solver.}
		\label{ConvergenceGraph}
	\end{center}
\end{figure}

\begin{figure}[!h]
	\begin{center}
		\includegraphics[width=\textwidth]{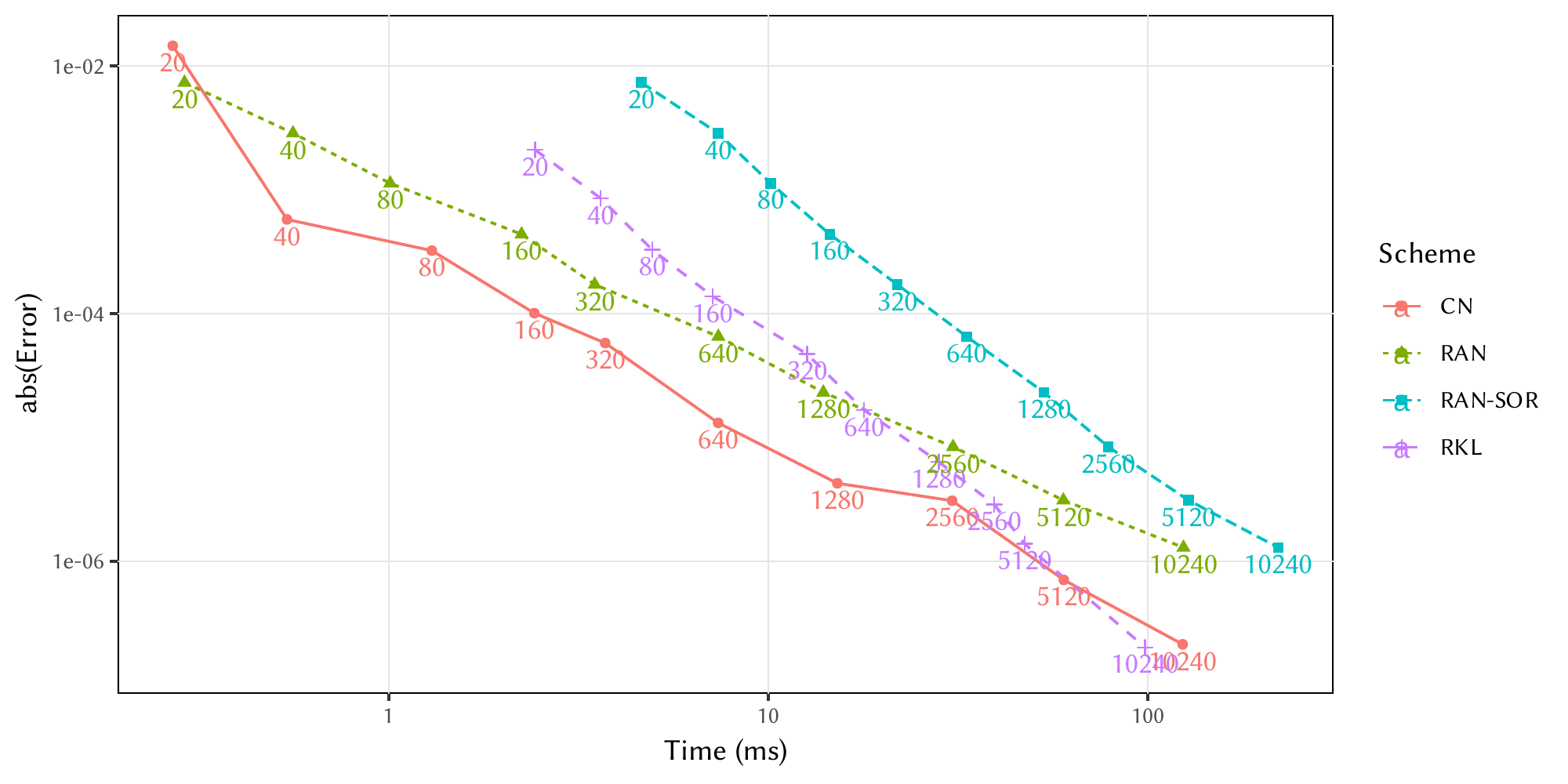}
		\caption{Error against wall-time to price an American Put with a fixed space step and increasing the number of time-steps. The labels on the points indicate the corresponding number of time-steps. Crank-Nicolson "CN", Rannacher "RAN" with the Brennan-Schwartz solver, "RAN-SOR" Rannacher  with the PSOR solver.}
		\label{ConvergenceGraphTime}
	\end{center}
\end{figure}

The RKL scheme requires 47 stages for 20 time-steps. It is then around 3 times slower than the Crank-Nicolson scheme, with the optimized tridiagonal policy iteration solver. For 640 time-steps, 9 stages are required, and the RKL scheme is then 1.5 times slower than Crank-Nicolson. The RKL scheme becomes faster only for more than 2560 time-steps, corresponding to 5 stages or less (Figure \ref{ConvergenceGraphTime}).

The convergence of the RKL scheme is as smooth as the Crank-Nicolson scheme with Rannacher time-stepping and noticeably more accurate. It follows closely the error profile of the pure Crank-Nicolson scheme when the number of time-steps is above 80 (Figure \ref{ConvergenceGraph}). The measured order of convergence in time, for a fixed number of space-steps, is 1.4 for all schemes considered here. In particular the schemes do not attain a second-order convergence in time on a fixed grid, likely because the early-exercise boundary moves in time, and does not fall on grid points.

\subsection{Order of convergence}
Like \citet{forsyth2002quadratic}, we look at the convergence on a successively refined grid in order to determine the global order of convergence. We use the same American option, with the same number of spacesteps and timesteps as \citet{forsyth2002quadratic}. The grid is uniform and varies in space from $x_0=0$ to $x_m=350$. An additional point is added such that the option strike falls exactly on a grid node. 

With a ratio of 4.0 or slightly higher between successive option value differences, the RKL scheme attains quadratic convergence. This is better than the Crank-Nicolson scheme with Rannacher time-stepping (Table \ref{table_forsyth_bs}). The latter exhibits an order of convergence slightly lower than 2, which mirrors the impact of the discontinuity in the second derivative of the option value across the exercise boundary on the Crank-Nicolson scheme. The Rannacher time-stepping only helps to smooth out the discontinuity in the initial condition, but not afterwards, along the exercise boundary.
	\begin{table}[htb]
\centering{
		\begin{tabular}{|c|c|r|r|r|r|r|}
			\hline Spacesteps & Timesteps & Value & Change & Ratio & Error & Time(s) \\ \hline		
			\multicolumn{7}{|c|}{Rannacher} \\ \hline
			68  & 25 & 14.633655 &  & & -4.52e-02 & 7.3e-05\\
			135 &50 & 14.665618 & 3.20e-02 & &-1.33e-02& 2.1e-04\\
			269 & 100 & 14.674848 & 9.23e-03 & 3.5 & 4.03e-03& 8.5e-04\\
			537 & 200 & 14.677598 & 2.75e-03 & 3.4 & -1.28e-03& 3.3e-03\\
			1073 & 400 & 14.678446 & 8.48e-04 & 3.2 & -4.32e-04 &8.8e-03\\ \hline
			\multicolumn{7}{|c|}{RKL} \\ \hline
			68 & 25 & 14.644920 &&& -3.40e-02& 1.0e-04\\
			135 & 50 & 14.670729 & 2.58e-02 & &-8.15e-03& 5.9e-04\\
			269 & 100 & 14.677086 & 6.36e-03 & 4.1 & -1.79e-03& 3.0e-03\\
			537 & 200 & 14.678572 & 1.49e-03 & 4.3 & -3.07e-04& 1.7e-02\\
			1073 & 400 & 14.678863 & 2.91e-04 & 5.1 & -1.54e-05 &1.0e-01\\ \hline
		\end{tabular} 
		\caption{Value of an American put option, $T=.25$, $r=.10$, $K=100$, $S=100$. Change is the difference in the solution from the coarser grid. Ratio is the ratio of the changes on successive grids. Constant timesteps. The strike is on the grid. The reference value is 14.67887836.}\label{table_forsyth_bs}
}
	\end{table}

As the grid becomes denser however, the number of stages used in the RKL scheme increases and the scheme becomes slower. The accuracy of the RKL scheme on a grid composed of 1073 space steps and 400 time-steps, measured against a reference value obtained on a very fine grid, is 1.54e-05. This is 28 times smaller than the accuracy of 4.32e-04, obtained by the Crank-Nicolson scheme with Rannacher time-stepping. On the grid composed of 537 space steps and 200 time-steps, the RKL scheme is slightly more accurate (3.07e-04). If we take into account the running time of both schemes, for a given accuracy, the two schemes exhibit a similar performance on this example.  


\section{Hamilton-Jacobi-Bellman PDE}
The uncertain volatility model developed in \citep{avellaneda1995pricing} is an example of (Hamilton-Jacobi-Bellman) HJB equation in finance. It supposes that the volatility is uncertain but bounded between a minimum volatility $\sigma_{\min}$ and a maximum volatility $\sigma_{\max}$. Under this model one can find the prices of a derivative product for a worst long strategy or a best long strategy. The optimal control problem is given by:
\begin{equation}
\frac{\partial f}{\partial t}(x,t) - \sup_{\sigma \in \hat{\sigma}} \lbrace \mu(x,t) x \frac{\partial f}{\partial x} + \frac{1}{2}\sigma^2 x^2\frac{\partial^2 f }{\partial x^2} - r(x,t)f(x,t) \rbrace = 0
\end{equation}
where $x$ is the underlying price, $\mu$ is the underlying drift, $\sigma$ its volatility and $r$ the interest rate, $\hat{\sigma}=\left[ \sigma_{\min}, \sigma_{\max} \right]$ and $F(x)=f(x,t_n)$ the option payoff at maturity.

In this model, the sign of the option $\Gamma$ will dictate the choice of $\sigma_{\min}$ or $\sigma_{\max}$ at each point of the grid. Even though the RKL scheme maintains the convex monotone property (CMP), the option greeks will not necessarily be smooth and this may impact the convergence of the scheme. As a consequence, this problem is particularly sensitive to oscillations in the scheme and the Crank-Nicolson scheme fails to converge to the correct solution without additional implicit time-steps to smooth-out the initial condition. 

Like \citep{pooley2003numerical} we price a butterfly spread option under the uncertain volatility model using the same parameters and using a uniform grid in space from $x_0=0$ to $x_m=150$ doubling the size in both the space dimension and the time dimension in order to look at the convergence. A butterfly spread corresponds to a long position in 2 calls at strike $K_1$ and $K_2$ and a short position in 2 calls at strike $\frac{K_1+K_2}{2}$. Our butterfly spread option is defined with the strikes $K_1=90$ and $K_2=110$ expiring at $T=0.25$ years. The interest rate is $r=10\%$, and the volatility is between $\sigma_{\min}=0.15$ and $\sigma_{\max}=0.25$. We solve the HJB via the same Newton method as in \citep{pooley2003numerical}. This means that we solve the non-linearity in the same Newton iteration for the $s$ stages of the RKL scheme. Only two iterations are necessary in practice.

Table \ref{table_uncertain} shows that in practice, the RKL scheme converges quadratically to the correct value, in contrast to the Crank-Nicolson scheme. The price given in \citep{pooley2003numerical} for the worst long strategy of the same option is 2.29769. To reach an accuracy of $10^{-3}$, the RKL scheme requires only 2.4 ms, while the backward Euler scheme requires a much finer grid of 1920 space-steps and 800 time-steps, and takes 100 ms to value the butterfly spread option. Although not displayed here, the Rannacher scheme requires around 3 ms and becomes more efficient on finer grids than the RKL scheme by a factor of two. The RKL scheme is however much simpler to implement.
	\begin{table}[thbp]
\tiny{\centering
		\begin{tabular}{r r r r r r r r r r r}\toprule
		 Spacesteps & Timesteps & 	\multicolumn{4}{c}{RKL ($\epsilon=0$)} &			\multicolumn{4}{c}{Backward Euler} \\  \cmidrule(lr){3-8}  \cmidrule(lr){7-10} 
		 & &		Value & Change & Ratio & Time & Value & Change & Ratio & Time\\\midrule
			60 & 25 & 2.305608 &  & 		& 0.1	 & 2.362065 &  &  &0.1 \\
			120 & 50 & 2.299416 & -6.19e-03& & 0.5 	 & 2.327263 & -3.48e-02 &  &0.4 \\
			240 & 100 & 2.298082 & -1.33e-03 & 4.6 & 2.1  & 2.312159 & -1.51e-02 & 2.3 &1.5\\			
			480 & 200 & 2.297787 & -2.95e-04 & 4.5 & 13.6 & 2.304852 & -7.31e-03 & 2.1 &6.9\\
			960 & 400 & 2.297710 & -7.68e-05 & 3.8  & 71.6 & 2.301251 &-3.60e-03 & 2.0 &24.2\\\bottomrule
		\end{tabular} 
		\caption{Value of a butterfly spread option. Change is the difference in the solution from the coarser grid. Ratio is the ratio of the changes on successive grids. Constant timesteps. The reference price is 2.29769. The valuation time is measured in ms.}\label{table_uncertain}
}
	\end{table}



\begin{figure}[h]
	\centering{
		\subfigure[\label{fig:gamma_uvm_digital960}$\epsilon=0$.]{
			\includegraphics[width=.45\textwidth]{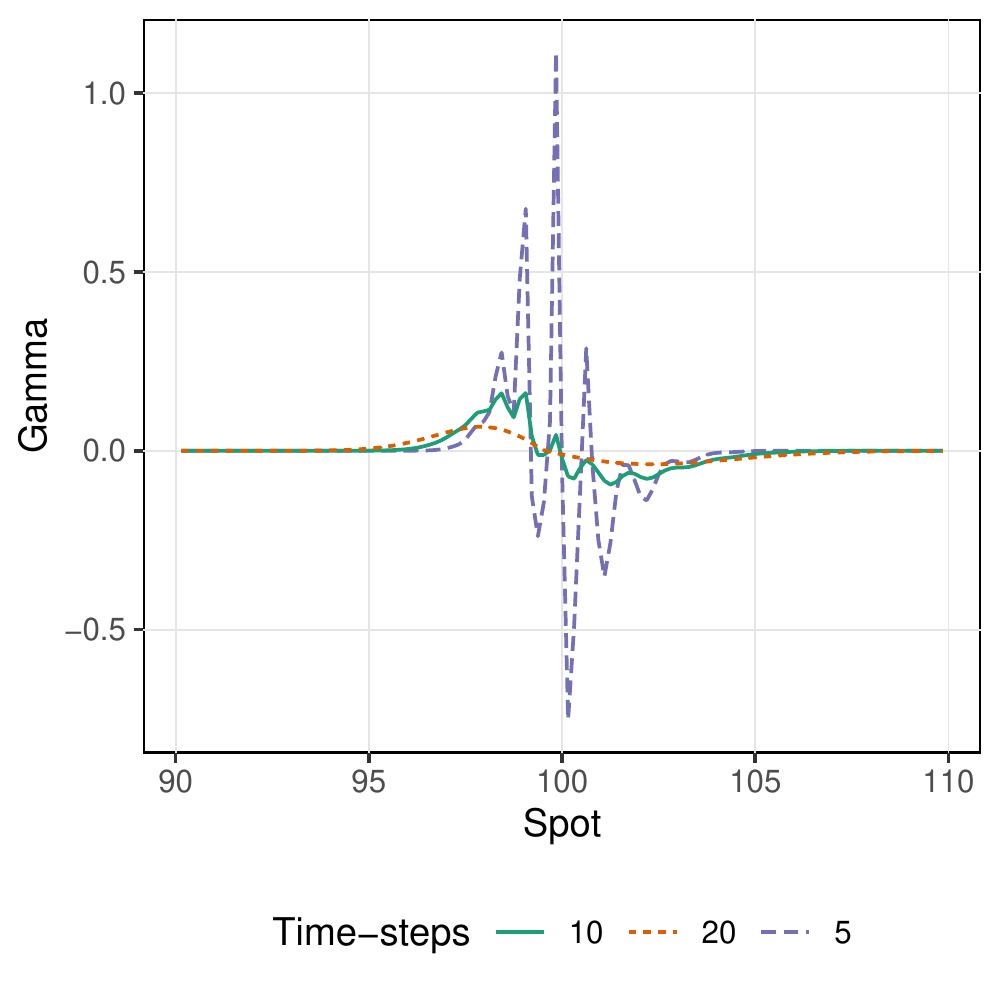}}
		\subfigure[\label{fig:gamma_uvm_digital960_e20}$\epsilon=20$.]{
			\includegraphics[width=.45\textwidth]{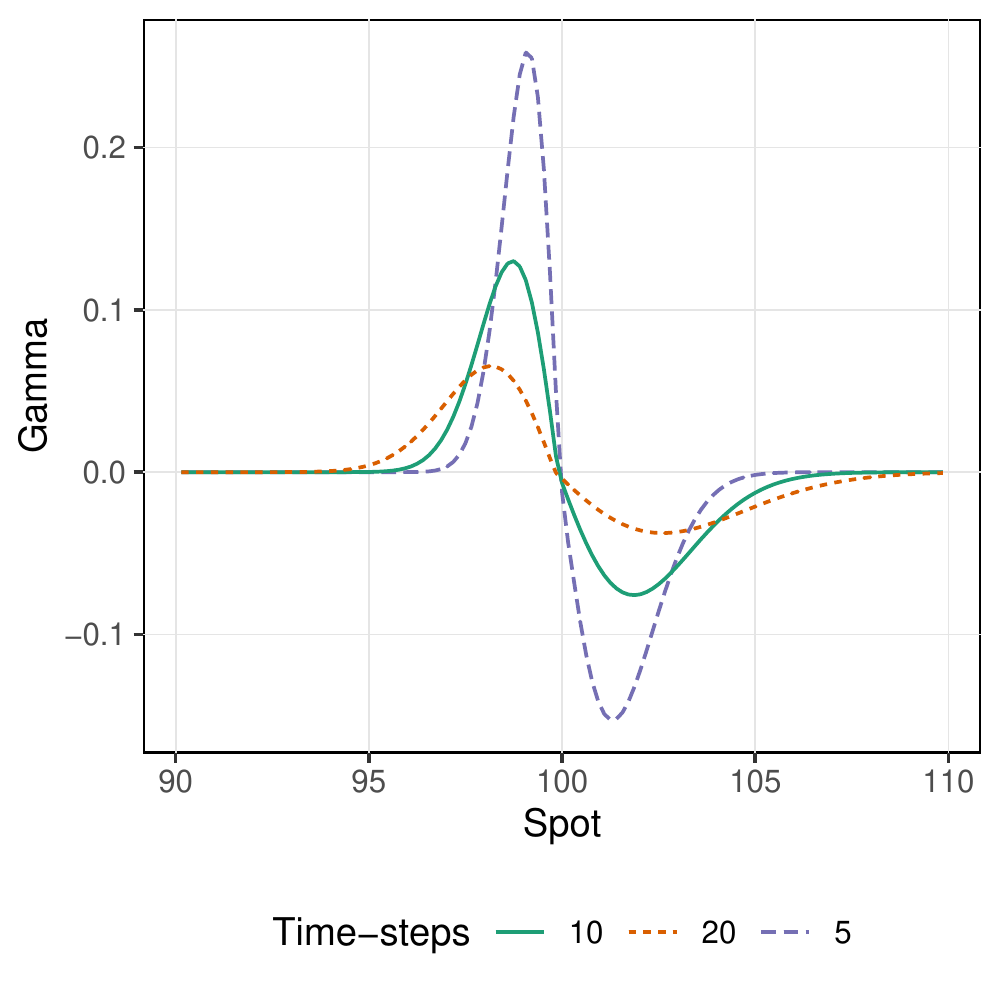} }
		\caption{Digital option $\Gamma$ in the uncertain volatility model obtained by the RKL scheme with and without shift, after 5, 10 and 20 time-steps on a grid composed of 400 time-steps and 960 space-steps.}}
\end{figure}

The case of a digital option of strike $K=100$, paying \$1 at maturity $T=0.25$ is more challenging, as the initial condition is discontinuous. We apply the second-order Kreiss smoothing on the initial condition \citep{kreiss1970smoothing}. Without shifting the Legendre polynomial, the convergence of the RKL scheme is erratic, because of internal oscillations, which impact the $\Gamma$ and thus the choice of volatility in the grid (Figure \ref{fig:gamma_uvm_digital960}). In particular, the oscillations are stronger compared to the butterfly spread option, due to the choice of initial condition. The default damping of the RKL scheme is not strong enough anymore to ensure a proper order of convergence.  

Figure \ref{fig:gamma_uvm_digital960_e20} shows that a large shift $\epsilon=20$ is necessary to increase damping and Table \ref{table_uncertain_digital} confirms that this shift restores proper convergence. 
The order of convergence is however just one on this example. In \citep{pooley2003numerical}, the Crank-Nicolson scheme with Rannacher time-stepping was also found to be only order-1 on this example. The timings are similar as for the case of the butterfly spread option (Table \ref{table_uncertain}) and the RKL scheme is almost four times slower than the backward Euler scheme on this finest grid considered.
\begin{table}[thbp]
	\small{\centering
		\begin{tabular}{r r r r r r r r r}\toprule
			Spacesteps & Timesteps & 	\multicolumn{3}{c}{RKL ($\epsilon=20$)} &			\multicolumn{3}{c}{Backward Euler} \\  \cmidrule(lr){3-5}  \cmidrule(lr){6-8} 
			& &		Value & Change & Ratio & Value & Change & Ratio\\\midrule
			60 & 25 & 0.459223 &  & 			 & 0.459283 &  &  \\
			120 & 50 & 0.450698 &-8.52e-03 &  	 & 0.450687 & -8.60e-03 &  \\
			240 & 100 & 0.446361 & -4.34e-03 & 2.0  & 0.446345 &-4.34e-03 & 2.0 \\
			
			480 & 200 & 0.444146 & -2.22e-03 & 2.0 &0.444137 &-2.21e-03 & 2.0 \\
			960 & 400 & 0.443018 & -1.13e-03 & 2.0  & 0.443014 & -1.12e-03 & 2.0 \\\bottomrule
		\end{tabular} 
		\caption{Value of a binary option. Change is the difference in the solution from the coarser grid. Ratio is the ratio of the changes on successive grids. Constant timesteps. The reference price is 0.441964}\label{table_uncertain_digital}
	}
\end{table}

Even though the first order backward Euler scheme appears preferable for the valuation of digital options under the uncertain volatility model, this example illustrates the importance of adding a shift to the RKL scheme to achieve convergence for some problems.

\section{Heston stochastic volatility model}
In the stochastic volatility model of \citet{heston93}, the asset $X$ follows
\begin{subequations}
	\begin{align}
	\diff X(t) &= (r(t)-q(t)) X(t)dt+\sqrt{V(t)} X(t) \diff W_X(t)\,,\label{eqn:heston_X}\\ 
	\diff V(t) &= \kappa \left(\theta - V(t)\right) + \sigma \sqrt{V(t)} \diff W_V(t)\,,\label{eqn:heston_V}
	\end{align}
\end{subequations}
with $W_X$ and $W_V$ being two Brownian motions with correlation $\rho$, and $r, q$ the instantaneous growth and dividend rates.

The corresponding PDE for the option price $f$ reads
\begin{equation}
\frac{\partial f}{\partial t} = \frac{v x^2}{2}\frac{\partial^2 f}{\partial x^2} + \rho\sigma x v \frac{\partial^2 f}{\partial x \partial v} + \frac{\sigma^2 v}{2} \frac{\partial^2 f}{\partial v^2} + (r-q)x\frac{\partial f}{\partial x} + \kappa(\theta - v)\frac{\partial f}{\partial v} - r f\,,\label{eqn:heston_pde}
\end{equation}
for $0 \leq t \leq T$, $x > 0$, $v > 0$, with initial condition  $f(T, x, v) = F(x)$.

\subsection{Heston PDE discretization}
The PDE will be solved on a truncated\footnote{Instead of truncation, a variable transformation, which converts the infinite domain to a finite domain may be more appropriate, as it would allow a global error control. The results and conclusions presented would however not change.} domain $[x_{\min}, x_{\max}]\times[v_{\min},v_{\max}]$.
We will typically truncate the $x$ domain to four standard deviations. A simple approximation is given by
\begin{align*}
x_{\min} = 0\,,&\quad x_{\max} =K e^{+4\sqrt{\theta T}}\,.
\end{align*}
where $K$ is the option strike price.
For the $v$ domain, let $\Phi_{\chi}(y,d_{\chi},\lambda_{\chi})$ be the cumulative distribution for the non-central chi-square distribution with $d_{\chi}$ degrees of freedom and non-centrality parameter $\lambda_{\chi}$. The distribution of the variance process $V(T)$ conditional on $V(0)$ is known \citep{cox1985theory,andersen2007efficient}, and thus we may choose
\begin{align*}
v_{\min} = 0\,,\quad v_{\max} = \Phi_{\chi}^{-1}(1-\epsilon_v, d_{\chi}, v_0 n_\chi) \frac{e^{-\kappa T}}{n_{\chi}}\,,
\end{align*}
with $d_{\chi} = 4\frac{\kappa\theta}{\sigma^2}$, $n_\chi = 4\kappa \frac{e^{-\kappa T}}{\sigma^2\left(1-e^{-\kappa T}\right)}$, and $\epsilon_v=10^{-4}$. 

At $v =v_{\max}$, we follow \citet[p. 385-386]{andersen2010interestcv} and let the price be linear in the variance dimension:
\begin{equation}\frac{\partial^2 f}{\partial v^2}(x,v,t) = \frac{\partial^2 f}{\partial x\partial v}(x,v,t) = 0\,,\label{eqn:vmin_boundary}\end{equation}
for $x \in (x_{\min},x_{\max})$. 
A priori, it is not guaranteed that Equation (\ref{eqn:vmin_boundary}) is a sufficiently accurate approximation at the lower boundary $v_{\min}$. When $v_{\min} = 0$, the exact boundary condition at $v=v_{\min}=0$ corresponds to the PDE obtained by setting $v=0$ and is preferable in practice. With the choice $v_{\min}=0$, the two conditions are equivalent.

At $x_{\max}$ and $x_{\min}$, we consider that the value is linear along $x$, which leads to
\begin{align}
\frac{\partial^2 f}{\partial x^2}(x,v,t) =\frac{\partial f}{\partial v}(x,v,t) = 0\,, \label{eqn:linear_bc_sv}
\end{align}
for  $v \in [v_{\min},v_{\max}]$  and $x \in \left\{x_{\min},x_{\max}\right\}$.

With a second-order central discretization of the derivatives, we obtain the following equation for the explicit step involved at each stage of the RKL scheme:
\begin{align*}
\hat{f}^\eta_{i,j} =& \hat{g}^{\eta-1}_{i,j} +\bar{\lambda}_\eta a_{i,j}\hat{f}^{\eta-1}_{i-1,j} + \bar{\lambda}_\eta b_{i,j} \hat{f}^{\eta-1}_{i,j} + \bar{\lambda}_\eta c_{i,j}\hat{f}^{\eta-1}_{i+1,j} + \bar{\lambda}_\eta d_{i,j}\hat{f}^{\eta-1}_{i,j-1} + \bar{\lambda}_\eta e_{i,j}\hat{f}^{\eta-1}_{i,j+1}\\
 &+\bar{\lambda}_\eta \omega_{i,j} \left(\hat{f}^{\eta-1}_{i+1,j+1} - \hat{f}^{\eta-1}_{i+1,j-1} - \hat{f}^{\eta-1}_{i-1,j+1} + \hat{f}^{\eta-1}_{i-1,j-1}\right)\,,
\end{align*}
with
\begin{subequations}
	\begin{align}
	a_{i,j} =& -\frac{k}{h_i\left(h_{i+1}+h_i\right)}\left(\mu_{i} h_{i+1} x_i - \beta^x_{i,j} v_j x_i^2\right) \label{eqn:stencil_eq1}\,,\\
	b_{i,j} =& - k \left(r_{i}+ \frac{\mu_{i}(h_i-h_{i+1})+\beta^x_{i,j} v_j x_i^2 }{h_i h_{i+1}} +  \frac{\kappa(\theta-v_j)(w_j-w_{j+1})+\beta^v_{i,j} v_j \sigma^2 }{w_j w_{j+1}}\right)\,, \label{eqn:stencil_eq2}\\
	c_{i,j} =& \frac{k }{h_{i+1}\left(h_{i+1}+h_i\right)}\left(\mu_{i}h_i x_i+ \beta^x_{i,j} v_j x_i^2\right) \label{eqn:stencil_eq3}\,,\\
	d_{i,j} =& -\frac{k  }{w_{j}\left(w_{j+1}+w_j\right)}\left(\kappa (\theta-v_j)w_{j+1} - \beta^v_{i,j} v_j \sigma^2\right) \label{eqn:stencil_eq4}\,,	\\
	e_{i,j} =& \frac{k  }{w_{j+1}\left(w_{j+1}+w_j\right)}\left(\kappa (\theta-v_j)w_j + \beta^v_{i,j} v_j \sigma^2\right) \label{eqn:stencil_eq5}\,,\\
	\omega_{i,j} =& \frac{k \rho\sigma  x_i v_j}{(h_i+h_{i+1})(w_j + w_{j+1})} \label{eqn:stencil_eq6}\,,
	\end{align}
\end{subequations}
and $h_i = x_i - x_{i-1}$, $w_j = v_j-v_{j+1}$, for $i=1,...,m-1$, and $j=1,...,n-1$. The classic central discretization correspond to the choice $\beta^x_{i,j}=\beta^v_{i,j}=1$.

In order to simplify the notation, we dropped the time index and all the coefficients $\mu_i, r_i, k$ involved are understood to be taken at a specific time-step. The index $j$ applies to the variance dimension instead of the time dimension in the previous sections of this paper.

The cell P\'eclet number $P$ is the ratio of the advection coefficient towards the diffusion coefficient in a cell \citep{hundsdorfer2013numerical}. When the P\'eclet condition $P \leq 2$ does not hold, the stability of the finite difference scheme is not guaranteed anymore: the solution may explode.  Here, the cell P\'eclet number for each dimension is 
\begin{align}
P^{x}_{i, j}(\beta^x_{i,j}) = \frac{2h_i}{\beta^x_{i,j}  v_j x_i} \left(r_i-q_i\right)&\,,\quad P^{v}_{i,j}(\beta^v_{i,j}) =\frac{2 w_j\kappa(\theta-v_j)}{\beta^v_{i,j} \sigma^2 v_j}\,.
\end{align}                                   
The P\'eclet conditions $P^{x}_{i,j} < 2$ and $P^{v}_{j} < 2$ do not necessary hold with typical values for the Heston parameters. This happens when $v_j$ is very small, which is generally the case for the first few indices $j$.  In order to ensure that the P\'eclet conditions hold, we will use the exponential fitting technique of \citet{allen1955relaxation, il1969differencing} when $P^{x}_{i,j} \geq 2$ as well as when $P^{v}_{i,j} \geq 2$. It consists in using the coefficients \begin{align}
\beta^{x}_{i,j} = \frac{P^{x}_{i,j}(1)}{2\tanh\left(\frac{P^{x}_{i,j}(1)}{2}\right)} &\,,\quad \beta^{v}_{i,j} = \frac{P^{v}_{i,j}(1)}{2\tanh\left(\frac{P^{v}_{i,j}(1)}{2}\right)}\,,\label{eqn:peclet}
\end{align}
instead of $\beta^{x}_{i,j}=\beta^{v}_{i,j}=1$. 

The boundary conditions, discretized with order-1 forward and backward differences, lead to
\begin{align*}
a_{i,0} &= -\frac{k}{h_i\left(h_{i+1}+h_i\right)}\left(\mu_{i} h_{i+1} x_i - \beta^x_{i,0} v_0 x_i^2\right)\,,\\
c_{i,0} &= \frac{k  }{h_{i+1}\left(h_{i+1}+h_i\right)}\left(\mu_{i}h_i x_i+ \beta^x_{i,0} v_j x_i^2\right) \,,\\
b_{i,0} &= - k \left(r_{i}+ \frac{\mu_{i}(h_i-h_{i+1})+\beta^x_{i,0} v_0 x_i^2 }{h_i h_{i+1}} +  \frac{\kappa(\theta-v_0)}{w_1}\right)\,,\\
d_{i,0} &= 0\,,	\quad e_{i,0} = \frac{k \kappa (\theta-v_0) }{w_{1}}\,,\quad w_{i,0} = 0\,,\\
a_{i,n} &= -\frac{k}{h_i\left(h_{i+1}+h_i\right)}\left(\mu_{i} h_{i+1} x_i - \beta^x_{i,n} v_n x_i^2\right)\,,\\
 c_{i,n} &= \frac{k  }{h_{i+1}\left(h_{i+1}+h_i\right)}\left(\mu_{i}h_i x_i+ \beta^x_{i,n} v_j x_i^2\right) \,,\\
b_{i,n} &= - k \left(r_{i}+ \frac{\mu_{i}(h_i-h_{i+1})+\beta^x_{i,n} v_n x_i^2 }{h_i h_{i+1}} -  \frac{\kappa(\theta-v_n)}{w_{n-1}}\right)\,,\\
d_{i,n} &= -\frac{k \kappa (\theta-v_n) }{w_{n-1}}\,,	\quad e_{i,n} = 0\,,\quad \omega_{i,n} = 0\,,
\end{align*}
for $i=1,...,m-1$, and
\begin{eqnarray*}
a_{0,j}=0\,, \quad {b}_{0,j} = - k\left(r_{0}+\frac{\mu_{0}x_0}{h_1}\right)\,,\quad  {c}_{0,j} = k \frac{\mu_{0}x_0}{h_1}\,,\quad d_{0,j}=e_{0,j}=\omega_{0,j}=0\,,\\
{a}_{m,j}  = - k\frac{\mu_{m}x_m}{h_{m}}\,,\quad  {b}_{m,j} = - k\left(r_{m}-\frac{\mu_{m}x_m}{h_{m}}\right)\,,\quad c_{m,j} = d_{m,j}=e_{m,j}=\omega_{m,j}=0\,,
\end{eqnarray*}
for $j=0,...,n$.

\subsection{American option price convergence}
\begin{figure}[h]
	\centering{
		\subfigure[\label{fig:heston_timeconvergence}Feller violated.]{
			\includegraphics[width=.45\textwidth]{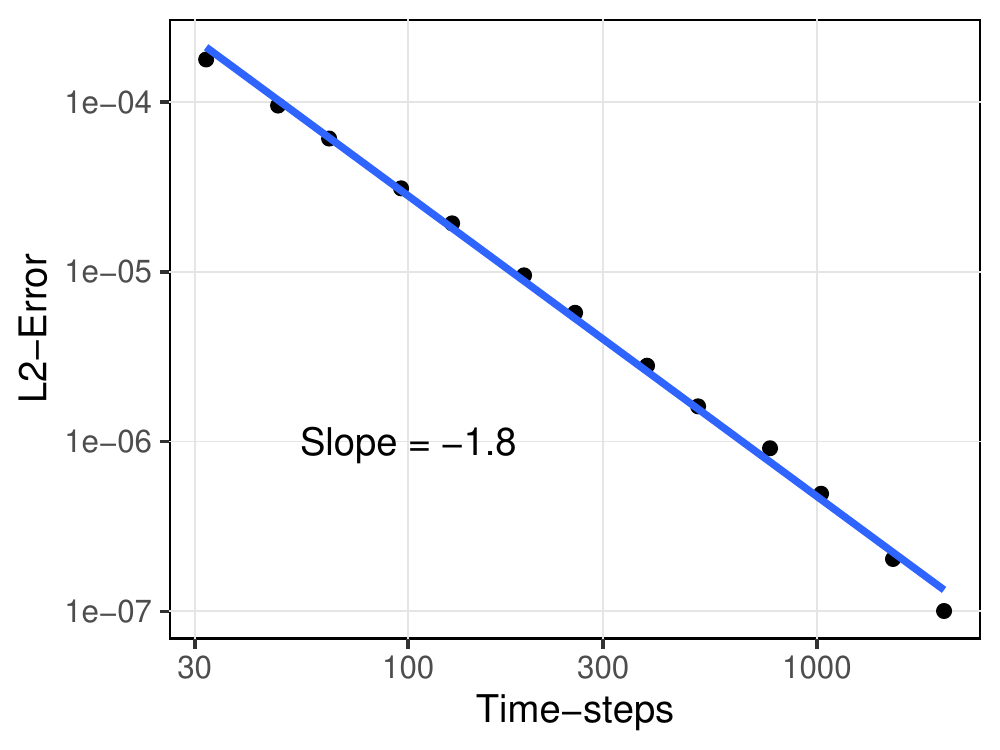}}\hspace{0.05\textwidth}
		\subfigure[\label{fig:heston_caseb_timeconvergence}Feller satisfied.]{
			\includegraphics[width=.45\textwidth]{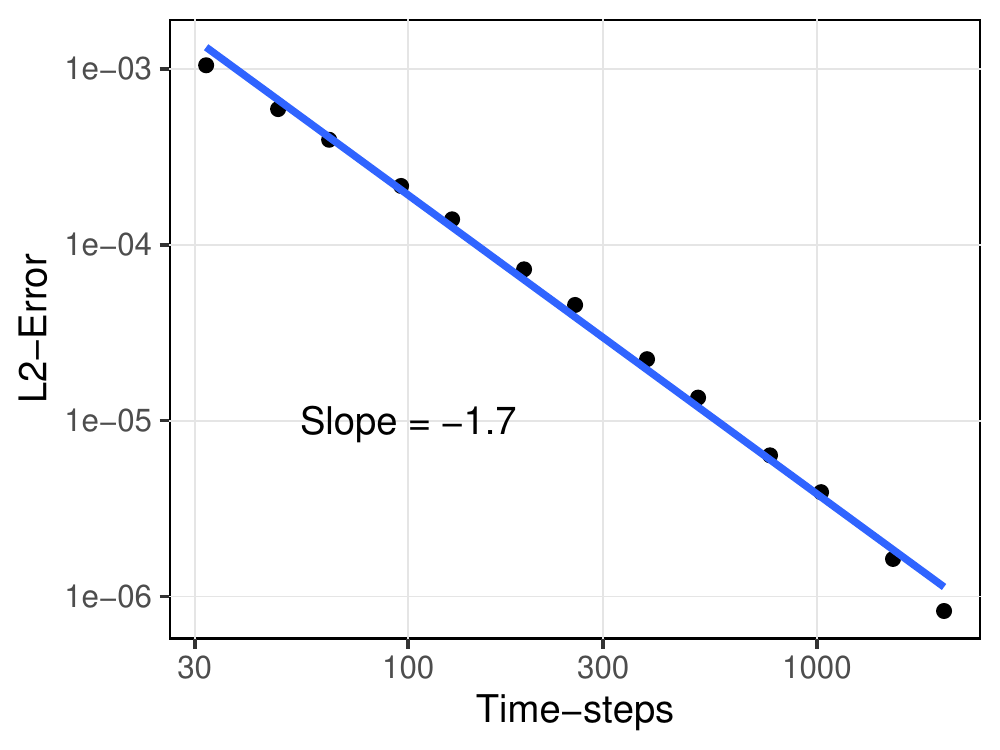} }
		\caption{$L_2$-error of the American put option price with the RKL scheme, varying the number of time-steps $l$, keeping the number of steps $m, n$ in the asset and variance dimensions fixed, for two different sets of Heston parameters. One set verifies the Feller condition and the other violates it.}}
\end{figure}

We first consider the case of an option of strike $K=100$ and maturity $T=0.25$, with Heston parameters $\kappa=1.15$, $\theta=0.0348$, $\sigma=0.39$, $\rho=-0.64$ and $r=0.04$, $q=0$ from \citet{fang2011fourier, haentjens2015adi}. Those are realistic values when fitting the Heston stochastic volatility model to market prices of equity options, and they violate the Feller condition.
We also consider the parameters $\kappa=0.6067$, $\theta=0.0707$, $\sigma=0.2928$, $\rho=-0.7571$ and $r=0.03$, $T=3$ from \cite{winkler2001valuation, haentjens2015adi}, which represent a longer maturity and do not violate the Feller condition.
Like \citet{haentjens2015adi}, we plot the error in the American option price for a fixed number of steps $m$ and $n$ in the asset and variance dimensions in Figures \ref{fig:heston_timeconvergence} and \ref{fig:heston_caseb_timeconvergence}. The reference price is obtained with a very large number of time-steps $l=4096$. We use a uniform grid composed of $m=100$ steps, $n=50$ steps, extending to three standard deviations, and smooth the payoff with method of \citet{kreiss1970smoothing}.
The order of convergence in time is measured by the slope of a linear least-square fit of the $L_2$-error in prices and is respectively of 1.7 and 1.8.

Table \ref{tbl:fang_prices} lists reference prices of the American put option of strike $K=100$ and maturity $T=0.25$ under the Heston model with parameters $\kappa=1.15$, $\theta=0.0348$, $\sigma=0.39$, $\rho=-0.64$ and $r=0.04$, $q=0$, a case where the Feller condition is violated.
\begin{table}[th]
	\tiny{\centering
		\begin{tabular}{l r r r}\toprule
			Method & \multicolumn{3}{c}{Price} \\\cmidrule(lr){2-4}
			& $S=90$ & $S=100$ & $S=110$ \\		
			COS from \citet{fang2011fourier} $N=60$ & 9.9958 & 3.2079 & 0.9280\\
			MCS-IT from \citet{haentjens2015adi} $(m,n,l)=(300,150,60)$ & 10.0039 & 3.2126 & 0.9305 \\
			\bottomrule
		\end{tabular} 
		\caption{American put option of strike price $K = 100$, using the settings of \citet{fang2011fourier}.\label{tbl:fang_prices}}
	}
\end{table}
Table \ref{tbl:fang_prices_rkl} shows the RKL scheme to be competitive with our implementation of the MCS-IT scheme of \citet{haentjens2015adi}. Both schemes use the exact same grid. Our results differ slightly from \citet{haentjens2015adi} because our non-uniform grid is set up differently. The RKL scheme is more efficient than the MCS-IT scheme on relatively sparse grids. On denser grids, the MCS-IT scheme becomes faster than the RKL scheme.
\begin{table}[th]
	{\centering
		\begin{tabular}{l r r r r r}\toprule
			$(m,n,l)$ & Method  & \multicolumn{3}{c}{Price} & Time (s) \\ \cmidrule(lr){3-5}
			&			& $S=90$ & $S=100$ & $S=110$ \\			\midrule
			$(75,38,120)$			& RKL & 10.0047& 3.2081 & 0.9302 & 0.03\\
			&	MCS-IT & 10.0050 & 3.2081 & 0.9301 & 0.11 \\
			$(150,75,240)$			& RKL & 10.0032 & 3.2088 & 0.9289 & 0.28\\
			&	MCS-IT & 10.0035 & 3.2088 & 0.9288 & 0.90 \\
			$(300,150,480)$	& RKL & 10.0027 & 3.2090 & 0.9285 & 3.12\\
			&	MCS-IT & 10.0028 & 3.2090 & 0.9285 & 5.94 \\
			$(300,150,60)$ &	RKL & 10.0028 & 3.2093 & 0.9287 & 0.96 \\
			&	MCS-IT & 10.0025 & 3.2090 & 0.9284 & 0.74 \\
			\bottomrule
		\end{tabular} 
		\caption{American put option of strike price $K = 100$, using the settings of \citet{fang2011fourier}. The CPU time in seconds is measured using Julia 1.5 on an Intel Core i7-7600U CPU.\label{tbl:fang_prices_rkl}}
	}
\end{table}

In Table \ref{tbl:heston_grid_convergence}, we consider an option of strike $K=10$ and maturity $T=0.25$, with Heston parameters $\kappa=5$, $\theta=0.16$, $\sigma=0.9$, $\rho=0.1$ and $r=0.1$, $q=0$, also studied in \citep{oosterlee2003multigrid,ikonen2008efficient,o2010pricing,mollapourasl2019radial}. Those values are less realistic, but allow a comparison with the existing literature. In order to compute the error, we use the reference values of \citet{ikonen2008efficient}, which may not be accurate up to the last digit, hence the slightly lower order of convergence on the finest grid. The grid is doubled in each dimension successively and a quadratic order of convergence corresponds  to a ratio of successive errors equal to 4.
	\begin{table}[thbp]
	{\centering
		\begin{tabular}{l r r r r r}\toprule
			Grid & \multicolumn{2}{c}{$v_0=0.0625$} &\multicolumn{2}{c}{$v_0=0.25$} & Time (s) \\\cmidrule(lr){2-3}\cmidrule(lr){4-5}
$(m,n,l)$&			$L_2$ error & Ratio & $L_2$ error & Ratio & \\\midrule
			(64, 32, 66) & 8.34E-4 & & 5.92E-4 & & 0.02 \\
			(128, 64, 130) & 1.89E-4 & 4.4 & 1.45E-4 & 4.1 & 0.14 \\
			(256, 128, 260) & 4.74E-5 & 4.0 & 3.60E-5 & 4.0 & 1.55 \\
			(512, 256, 514) & 1.35E-5 & 3.5 &9.42E-6& 3.8 & 23.27 \\\bottomrule
		\end{tabular} 
		\caption{Errors calculated using five American put prices with asset spot prices of
$x = 8, 9, ..., 12$ with Heston parameters  $\kappa=5$, $\theta=0.16$, $\sigma=0.9$, $\rho=0.1$ and $r=0.1$, $q=0$. Also reported are the
			the ratio of consecutive errors, and the CPU times in seconds using Julia 1.5 on an Intel Core i7-7600U CPU.\label{tbl:heston_grid_convergence}}
	}
\end{table}
The grids correspond exactly to the settings of \citet{ikonen2008efficient} and the number of time-steps chosen is conservative. A similar accuracy is reached for the first two grids with $l=8$ and $l=16$.	
Per Table \ref{tbl:heston_grid_convergence}, the measured order of convergence is around 2 and the RKL scheme is competitive with the component-wise splitting of \citet{ikonen2008efficient}, which is much more involved in terms of implementation.

\section{Conclusion}
We have described how to apply the Runge-Kutta-Legendre (RKL) scheme to price American options under the Black-Scholes, local volatility, uncertain volatility, and Heston stochastic volatility models.

The main benefit of RKL is to be a second-order explicit scheme. Being explicit, it is straightforward to implement. It also simplifies the handling of non-linearities: for the American option problem, the constraint is trivial to apply, while for an implicit scheme, a more involved specialized solver, such as the policy iteration solver of \citet{reisinger2012use}, or a more generic, slower solver such as the projective successive over relaxation \citep{WiDeHo93}, need to be used.
When the number of dimensions of the problem increases, implicit schemes require advanced splitting techniques to stay efficient. In contrast, the RKL scheme only involves matrix vector products, which are easy to parallelize on many cores.

Like Runge-Kutta-Chebyshev schemes, the number of internal stages can be increased to overcome any limitation of  the time-step size to ensure the absolute stability property.

Compared to Runge-Kutta-Chebyshev schemes, the RKL scheme stays stable without shift, and maintains the convex monotone property. This leads to a quadratic order of convergence on the problem of pricing American options. We have however shown that a shift, which increases the damping at each time-step, may be necessary in order to obtain smoother greeks, for example in the case of a low number of time-steps, or when the non-linearity depends on the greeks, such as in the case of the uncertain volatility model.

Compared to implicit schemes, the RKL scheme may, in some situations, require too many stages for stability, for example when the spectral radius of the explicit scheme discretization matrix is very large, which typically happens for a  step size very small in the asset dimension and relatively large in the time dimension in a non-uniform grid. While the RKL scheme is always much more efficient than the explicit Euler scheme, it may still be significantly slower than an implicit scheme in those situations.

Further research could explore the possibility of constructing higher-order methods based on the family of RKL schemes, along the lines of the fourth order ROCK methods based on Runge-Kutta-Chebyshev methods \citep{abdulle2002fourth}.

\bibliographystyle{ws-ijtaf}
\section*{Acknowledgments}The author would like to thank the anonymous referees for their precise comments, thus helping to improve the quality of this paper.

\appendix
\section{Reference values}
\subsection{Black-Scholes model}
Table \ref{tbl:osullivan_prices} lists the American put option prices corresponding to the example in \citep{OSullivan09}: strike price $K = 100$, maturity $T=1$, spot price $S = 100$, discount rate $r= 5\%$, and volatility $\sigma=20\%$. The space step size is 1.0 on a grid ranging from $S_{\min}=0$ to $S_{\max}=500$. This corresponds to 500 space steps.

	\begin{table}[thbp]
	{\centering
		\begin{tabular}{l l r r r}\toprule
			Time-steps & Scheme & Price & Error & Time(ms) \\\midrule
20 & CN & 6.07288219 &-1.46e-02 & 0.3\\
 & RAN& 6.08012467 &-7.37e-03 & 0.3\\
 & RAN-SOR& 6.08012467 &-7.37e-03 & 4.6\\
 & RKL&  6.08960614&  2.11e-03 & 2.3\\
40 & CN & 6.08691599 &-5.77e-04 & 0.5\\
 & RAN& 6.08462518 &-2.87e-03 & 0.5\\
 & RAN-SOR& 6.08462518 &-2.87e-03 & 7.0\\
 & RKL&  6.08834823&  8.55e-04 & 3.4\\
80 & CN &6.08717054  &-3.23e-04 & 1.0\\
 & RAN& 6.08635966 &-1.13e-03 & 1.0\\
 & RAN-SOR& 6.08635966 &-1.13e-03 & 9.1\\
 & RKL&  6.08782107&  3.28e-04 & 4.6\\
160 & CN& 6.08739244 &-1.01e-04 & 1.8\\
 & RAN& 6.08705508& -4.38e-04 & 1.8\\
 & RAN-SOR& 6.08705508& -4.38e-04 & 13.6\\
 & RKL&  6.08763115&  1.38e-04 & 6.8\\
320 & CN &6.08743541 &-5.79e-05 & 3.5\\
 & RAN& 6.08732094 &-1.72e-04 & 3.4\\
 & RAN-SOR& 6.08732094 &-1.72e-04 & 21.8\\
 & RKL&  6.08754051&  4.72e-05 & 11.2 \\
640 &CN &6.08748020 &-1.31e-05 & 7.2\\
 &RAN &6.08742806 &-6.53e-05 & 7.5\\
 &RAN-SOR &6.08742806 &-6.53e-05 & 33.0\\
 &RKL & 6.08750997 & 1.66e-05 & 15.1\\
1280 &CN & 6.08748906 & -4.26e-06 & 13.6\\
& RAN & 6.08747018 &-2.31e-05 & 14.0\\
& RAN-SOR & 6.08747018 &-2.31e-05 & 50.6\\
& RKL & 6.08749966 &6.34e-06 & 25.1\\ \bottomrule
		\end{tabular} 
		\caption{American put option of strike price $K = 100$, using the settings of \citet{OSullivan09}. CN, RAN, RKL stand for the Crank-Nicolson, Rannacher schemes with the Brennan-Schwartz tridiagonal solver, RAN-SOR for the Rannacher scheme with the PSOR solver, and RKL for the Runge-Kutta-Legendre scheme.\label{tbl:osullivan_prices}}
	}
\end{table}

\end{document}